# Polarizabilities of Rn-like Th$^{4+}$ from rf spectroscopy of Th$^{3+}$ Rydberg levels

Julie A Keele and S.R. Lundeen
*Department of Physics, Colorado State University, Fort Collins, Colorado 80523, USA*

C.W. Fehrenbach
*J.R. Macdonald Laboratory, Kansas State University, Manhattan, Kansas 66506, USA*

High resolution studies of the fine structure pattern in high-L n=37 levels of Th$^{3+}$ have been carried out using radio-frequency (rf) spectroscopy detected with Resonant Excitation Stark Ionization Spectroscopy (RESIS). Intervals separating L=9 to L=15 levels have been measured, and the results analyzed with the long-range effective potential model. The dipole polarizability of Th$^{4+}$ is determined to be $\alpha_D$= 7.720(7) a.u.. The quadrupole polarizability is found to be 21.5(3.9) a.u. Both measurements represent significant tests of *a-priori* theoretical descriptions of this highly relativistic ion.

## I. INTRODUCTION

The Radon-like Th$^{4+}$ ion is the most common charge state of thorium in chemical compounds. Understanding the chemistry of thorium and other actinide elements is important to many national priorities, but very little experimental data exists to check the *a-priori* theoretical methods used to describe the ions that are at the heart of that chemistry. For example, no optical spectroscopy of excited levels of Th$^{4+}$ has been reported. The most significant dynamic characteristics of the $^1S_0$ ground state of Th$^{4+}$ are its dipole and quadrupole polarizabilities. They control important aspects of the long-range interactions of the Th$^{4+}$ ion with ligands or electrons. These properties can be measured by studying the binding energies of high-*L* Rydberg levels of Th$^{3+}$, bound to the Th$^{4+}$ ground state. A recent study of this type using the RESIS technique reported the first measurements of the dipole and quadrupole polarizabilities of $Th^{4+}$ [1]. The reported precision was approximately 1% and 20% respectively. In order to obtain better measurements, we report here a new study in which rf spectroscopy was used to directly measure intervals separating n=37 levels of $Th^{3+}$ with $9 \leq L \leq 15$. This builds on the previous study since the RESIS method was used to detect the population change in particular high-$L$ n=37 levels caused by an rf-induced transition. The resulting measurement improves the precision of the dipole polarizability by an order of magnitude, but indicates that the previous report of the quadrupole polarizability was an over-estimate.

## II. EXPERIMENT

The technique used for this study is similar in principle to that used for a previous study of high-L Rydberg states in $Si^{2+}$ [2]. The apparatus used here is a slightly modified version of that used for two recent optical RESIS studies [1,3] and is illustrated schematically in Fig. 1. A beam of 100 keV $Th^{4+}$ ions is extracted from an electron cyclotron resonance (ECR) ion source, mass selected and focused through a dense Rb Rydberg target. A significant fraction (3-6%) of the $Th^{4+}$ ions capture a single electron from the target Rb atoms to form highly excited Rydberg levels of $Th^{3+}$. After charge capture, the $Th^{3+}$ beam is magnetically selected and focused by the strong electric field in an electrostatic lens, which also has the effect of Stark ionizing the very weakly bound $Th^{3+}$ levels that will later form the upper state of the RESIS excitation.

The Rydberg target was excited to the 10F level, so it is expected that the most populated $Th^{3+}$ levels have binding energies of approximately $0.3 \pm .1$ eV, i.e. be near $n=30$ [4]. Several possible RESIS transitions upwards from these levels were observed in the optical RESIS study [1]. One of them, the $n=37$ to $n'=73$ transition using the 10P(10) $CO_2$ laser line, was used for this study. As illustrated in Fig. 1, two $CO_2$ laser interactions regions were used here, separated by a region where an rf electric field could be applied. In both laser regions, the angle of intersection with the $Th^{3+}$ beam was adjusted to Doppler-tune the $CO_2$ laser into resonance with the excitation of one particular $n=37$ level. As an example, consider observation of the $L=11$ to $L=12$ transition, where both laser regions are tuned to excite the $n=37$, $L=11$ to $n=73$, $L=12$ transition. In this case, the first region depletes the population of the $L=11$ level, insuring a population difference with the neighboring $L=12$ level. The second laser region excites about half of the population in the $L=11$ level up to $n=73$ where it is subsequently Stark ionized. This creates a current of $Th^{4+}$ ions that can be charge selected and collected in a channel electron multiplier. If the frequency and amplitude of the rf electric field encountered between the laser regions is correct to drive a resonant transition between $L=11$ and $L=12$, this will change the population of $L=11$ entering the second laser region and the eventual current of $Th^{4+}$ ions collected after Stark ionization. To measure the transition, both lasers are left unchopped and the rf power is modulated at 2 kHz as a synchronous change in the $Th^{4+}$ current is measured. Figure 2 illustrates one such resonance curve observed as the frequency of the rf electric field is varied. The width of the resonance, approximately 2.4 MHz, is due to the transit time through the rf region and the unresolved splitting (1.03 MHz) between the two spin components of the transition. The Gaussian fit shown in Fig. 2 determined the center frequency to a precision of approximately 0.1 MHz.

The rf interaction region used here was a shorter version of the eccentric coaxial transmission line used in reference [2]. The active length of the region, 12.7 cm, produced an interaction time of 0.44 μs for the 100 keV $Th^{3+}$ ions. The diameters of the outer and inner conductors were 3.29 cm and 0.62 cm respectively, and the eccentric offset was chosen to produce an impedance of 50Ω [5]. The rf electric field propagated either parallel or anti-parallel to the ion beam velocity, and data was taken in both directions and averaged to obtain the unshifted resonance position. Table I lists the transitions measured in this work. Four single-photon transitions were measured. Since $L$=12 was the highest $L$ level that could be reliably resolved in the RESIS excitation, levels above $L$=13 were studied using two and three photon transitions [6]. At the higher rf powers needed for the multi-photon transitions, AC shifts of the resonance positions were significant. The shift rate of the 3-photon 12-15 transition was directly measured, and the quoted result in Table I represents the linearly extrapolated result at zero power. The much smaller AC shift of the 2-photon 12-14 transition was inferred from a calculation of the relative shift rates of the two and three photon transitions. The final column in Table I summarizes the measured fine structure intervals.

In earlier studies, one concern was the possible build up of stray electric fields within the rf interaction region that could Stark shift the transition under study. As a check on such effects in this study, the transition most sensitive to such shifts, the 3-photon 12-15 transition, was measured repeatedly during the data collection period. Comparison of the observed resonance positions gave no evidence of variation in the possible stray field. Once the pattern of fine structure intervals was measured, the presence of any stray field should be revealed, as in previous studies [2], by the effect of such a field on the pattern of fine structure intervals. The Stark shift rates of each interval are easily estimated once an approximate level pattern is known. Table II lists the calculated rates for each interval. In this case, a fit of the data pattern, including the effect of an arbitrary constant stray electric field, found a stray electric field amplitude consistent with zero with 0.01 V/cm precision. It thus appears that stray fields are not a significant factor in this measurement.

### III. ANALYSIS

The fine structure pattern in these high-$L$ Rydberg levels is primarily determined by the expectation value of an effective potential whose first two terms are:

$$V_{eff}(r) = -\frac{\alpha_D}{2}\frac{1}{r^4} - \frac{1}{2}(\alpha_Q - 6\beta_D)\frac{1}{r^6} + \ldots \quad (1)$$

The parameters $\alpha_D$ and $\alpha_Q$ are the adiabatic dipole and quadrupole polarizabilities and $\beta_D$ is the first non-adiabatic dipole polarizability. The difference in the expectation values of $V_{eff}$ in the two fine structure levels contributes most of the measured interval, but two small additional contributions come from a) relativistic corrections, and b) the second order effect of $V_{eff}$. The first is the standard first relativistic correction to the energy of a hydrogenic level. The second of these is calculated analytically in the approximation that only the first term in $V_{eff}$ is significant [7]. The full fine structure interval is expected to be given by:

$$E(n, L') - E(n, L) = C_4 \Delta\langle r^{-4}\rangle + C_6 \Delta\langle r^{-6}\rangle + \Delta E_{rel} + \Delta E^{[2]} \quad (2)$$

where $C_4$ and $C_6$ are the coefficients of $r^{-4}$ and $r^{-6}$ occurring in $V_{eff}$ containing the desired core properties. In order to extract these from the measured intervals, the last two contributions were calculated and subtracted, leaving modified intervals due only to the expectation value of $V_{eff}$, which we denote as $\Delta E^{[1]}$. Table II shows these corrections and the inferred values of $\Delta E^{[1]}$.

If Eq. (1) gives the only significant terms in $V_{eff}$, then the intervals $\Delta E^{[1]}$ are expected to form a straight line when scaled and plotted as suggested in Eq. 3.

$$\frac{\Delta E^{[1]}}{\Delta \langle r^{-4} \rangle} = B_4 + B_6 \frac{\Delta \langle r^{-6} \rangle}{\Delta \langle r^{-4} \rangle} \qquad (3)$$

Where the fitted coefficients $B_4$ and $B_6$ would be identical to the coefficients in $V_{eff}$, $C_4$ and $C_6$. The radial expectation values needed to form this scaled plot are hydrogenic values, scaled to account for the mass of the $Th^{4+}$ core ion. Fig. 3 shows such a scaled plot with a linear fit to Eq. 3 represented by the dotted line. Quite obviously, the data is not consistent with a simple straight line. Instead, it clearly shows curvature, suggesting that additional terms in $V_{eff}$, proportional to higher inverse powers of r are contributing to the measured intervals. Many of these terms have been formally calculated, giving an extended potential of the form

$$V_{eff}(r) = -\frac{\alpha_D}{2}\frac{1}{r^4} - \frac{1}{2}(\alpha_Q - 6\beta_D)\frac{1}{r^6} + \left(\frac{8Q}{5}\gamma_D + \frac{\delta}{2}\right)\frac{1}{r^7}$$

$$-\frac{18}{5}\gamma_D \frac{L(L+1)}{r^8} - C_8 \frac{1}{r^8} + .... \qquad (4)$$

where the coefficients occurring in $V_{eff}$ are all properties of the $Th^{4+}$ ion [8]. The parameter $\gamma_D$ is the second non-adiabatic dipole polarizability, and the parameter $\delta$ determines the size of the first adiabatic contribution from the third-order perturbation energy. The parameter $C_8$ represents the net contribution of several higher order terms. All of these parameters except $C_8$ are defined in terms of $Th^{4+}$ matrix elements and excitation energies in Appendix A. For simplicity, we will refer to the total coefficient of the several powers of r according to

$$V_{eff} = -\frac{C_4}{r^4} - \frac{C_6}{r^6} - \frac{C_7}{r^7} - C_{8L}\frac{L(L+1)}{r^8} - \frac{C_8}{r^8} + .... \quad (5)$$

Distinguishing the contributions of these various terms to fine structure energies depends on the fact that the expectation values of the higher inverse powers decreases more rapidly with L than that of the lower inverse powers. For that reason, the contribution proportional to $L(L+1)r^{-8}$ is considered separately from the term proportional to $r^{-8}$ since its behavior with L is similar to $r^{-7}$. The contributions of these several terms to a measured fine structure pattern was discussed in a previous study [8]. That study concluded that for a wide range of parameters $C_4$, $C_6$, $C_7$, $C_{8L}$, and $C_8$, the fine structure pattern could be fit with precision by including only terms proportional to $r^{-4}$, $r^{-6}$, and $r^{-8}$. In other words, it appears to be impractical to extract the five $C_i$ coefficients independently from a fit of the data such as represented in Fig. 3. Instead, the data can be accurately parameterized by three parameters, $B_4$, $B_6$, and $B_8$ in the function below.

$$\frac{\Delta E^{[1]}}{\Delta\langle r^{-4}\rangle} = B_4 + B_6 \frac{\Delta\langle r^{-6}\rangle}{\Delta\langle r^{-4}\rangle} + B_8 \frac{\Delta\langle r^{-8}\rangle}{\Delta\langle r^{-4}\rangle} \quad (6)$$

The solid line in Fig. 3 shows a fit to Eq. 6, an excellent fit to all the data. The fitted parameters are:

$$B_4 = 3.8590(33)$$
$$B_6 = 3.0(1.5)$$
$$B_8 = 391(91)$$

Because of the possible effects of the terms proportional to $C_7$ and $C_{8L}$, the interpretation of the fitted $B_4$ and $B_6$ parameters has to be done with caution. Since the fine structure energies are found to be accurately parameterized by $B_4$, $B_6$, and $B_8$ regardless of the relative contributions of the various term, each fitted parameter should be properly interpreted as arising from the contributions of all five terms. The terms proportional to $C_4$, $C_6$, and $C_8$ will contribute directly to the coefficients $B_4$, $B_6$, and $B_8$, but the terms proportional to $C_7$ and $C_{8L}$ may contribute to several of the fitted parameters. Following the treatment in reference [8], the influence of possible $C_7$ and $C_{8L}$ terms on the fit of the data in Fig. 3 was evaluated by plotting their contributions to the seven measured fine structure intervals for unit coefficient, scaling the contribution in the same way as the data is scaled to $\Delta\langle r^{-4}\rangle$, and fitting to Eq. 6, using the same relative weights as was used for the experimental measurements. An excellent quality fit was obtained in each case with the results;

$$\frac{\Delta\langle r^{-7}\rangle}{\Delta\langle r^{-4}\rangle} \rightarrow B_4 = -4.7(5) \times 10^{-5},\ B_6 = 0.0555(18),\ B_8 = 4.86(15)$$

$$\frac{\Delta\left(L(L+1)\langle r^{-8}\rangle\right)}{\Delta\langle r^{-4}\rangle} \rightarrow B_4 = -3.5(4) \times 10^{-4},\ B_6 = 0.397(13),\ B_8 = 44.4(1.1)$$

Where the uncertainties attached to the fitted coefficients reflect the precision of the fits. Based on these fits, it can be seen that non-zero values of $C_7$ and $C_{8L}$ would affect the fitted values of both $B_4$ and $B_6$ according to:

$$\begin{aligned} B_4 &= C_4 - [4.7(5) \times 10^{-5}]C_7 - [3.5(4) \times 10^{-4}]C_{8L} \\ B_6 &= C_6 + [0.056(2)]C_7 + [0.40(1)]C_{8L} \end{aligned} \quad (7)$$

and interpretation of the parameters that fit the data must take this into account. Rewriting Eq. 7 in term of core properties.

$$B_4 = \frac{\alpha_D}{2} - [1.0(1) x 10^{-3}]\gamma_D + [2.4(3) x 10^{-5}]\delta$$
$$B_6 = \frac{\alpha_Q}{2} - 3\beta_D + [1.07(5)]\gamma_D - [0.028(1)]\delta$$
(8)

Clearly, if both $\gamma_D$ and $\delta$ are zero, the simplest interpretation of the fit parameters in terms of core properties is valid, but if $\gamma_D$ or $\delta$ is sufficiently large, that interpretation may be significantly in error. Since the dependence of $B_4$ on these parameters is very slight, the most significant effect is likely to be in the parameter $B_6$.

In order to determine the significance of the $C_7$ and $C_{8L}$ contributions it is necessary to estimate the core properties $\gamma_D$ and $\delta$. This is relatively easy in the case of $\gamma_D$ since it is related to the same set of matrix elements and excitation energies that determine $\alpha_D$ and $\beta_D$.

$$\alpha_D \equiv \frac{2}{3}\sum_{\lambda'}\frac{\langle gJ=0\|\vec{D}\|\lambda'J=1\rangle^2}{\Delta E(\lambda')}, \quad \beta_D \equiv \frac{1}{3}\sum_{\lambda'}\frac{\langle gJ=0\|\vec{D}\|\lambda'J=1\rangle^2}{\Delta E(\lambda')^2}, \quad \gamma_D \equiv \frac{1}{6}\sum_{\lambda'}\frac{\langle gJ=0\|\vec{D}\|\lambda'J=1\rangle^2}{\Delta E(\lambda')^3}$$
(9)

Theoretical calculations [9,10] of the first two parameters, $\alpha_D$ = 7.75 a.u. and $\beta_D$ = 2.97 a.u. suggest that the average excitation energy of the contributing states is 1.3 a.u, and therefore that $\gamma_D \sim 1.1$ a.u. This value, which is probably good to 10% makes its contribution to $B_4$ smaller than the experimental error in this study. Its contribution to $B_6$, however, is significant.

Obtaining an estimate of the parameter $\delta$ is much more difficult. As the definition shows, $\delta$ is related to dipole and quadrupole matrix elements between core and excited levels, and the excitation energies of these levels.

$$\delta \equiv \frac{4\sqrt{2}}{15} \sum_{\lambda',\lambda''} \frac{\langle gJ=0\|\vec{D}\|\lambda'J=1\rangle\langle \lambda'J=1\|\vec{D}\|\lambda''J=2\rangle\langle \lambda''J=2\|\vec{\vec{Q}}\|gJ=0\rangle}{\Delta E(\lambda')\Delta E(\lambda'')}$$
$$+ \frac{2\sqrt{30}}{45} \sum_{\lambda',\lambda''} \frac{\langle gJ=0\|\vec{D}\|\lambda'J=1\rangle\langle \lambda'J=1\|\vec{\vec{Q}}\|\lambda''J=1\rangle\langle \lambda''J=1\|\vec{D}\|gJ=0\rangle}{\Delta E(\lambda')\Delta E(\lambda'')} \quad (10)$$

In view of the complete absence of experimental information about the position of such levels, this is a challenging undertaking. Fortunately, all the excited levels of $Th^{4+}$ are expected to be rather high in energy, making the denominators of Eq. 10 relatively large. Estimates of the level energies in the $6p^5 5f$, $6p^5 6d$ and $6p^5 7s$ configurations obtained in the Dirac Hartree Fock (DHF) approximation, and order of magnitude estimates of the relevant dipole and quadrupole matrix elements [11] suggest that the magnitude of $\delta$ does not exceed 30 a.u.. If this is true, then the contribution of $\delta$ to $B_4$ is negligible at the level of experimental precision in this study, but again the contribution to $B_6$ could be significant. Since the sign of $\delta$ is not necessarily positive or negative, we assign a value

$$\delta = 0 \pm 30 \quad a.u.$$

The fitted coefficient $B_4$ gives a measurement of $\alpha_D$.

$$\alpha_d = 2B_4 + 0.0020(2)\gamma_D = 7.720(7)$$

This result improves upon the precision of the result obtained from the optical data [1] by about an order of magnitude. Both values are shown in Table III, along with several theoretical calculations. Agreement to better than 0.5% is found both with a recent relativistic coupled-cluster (RCCSD(T)) calculation [12], and with a relativistic random phase approximation (RRPA) calculation [9]. A fully relativistic Dirac Hartree Fock calculation is in error by 16% and a non-relativistic Hartree Fock calculation misses by 33%.

The fitted value of $B_6$ can be related to the quadrupole polarizability of $Th^{4+}$, but according to Eq. 8 this is only one of four contributing factors to $B_6$. Using calculated estimates of $\beta_D$ ($\beta_D$ = 2.97 a.u.[10]), the simple estimate of $\gamma_D$ discussed above ($\gamma_D$ = 1.1), and the primitive estimate of $\delta$ ($\delta$=0 ± 30 a.u), the value of $\alpha_Q$ can be estimated as:

$$\alpha_Q = 2B_6 + 6\beta_D - 2.14\gamma_D + .056\delta$$
$$= 6.0(3.0) + 17.8(1.8) - 2.3(2) + 0(1.7) = 21.5(3.9)$$

Clearly, this is not a purely experimental result, since it depends on calculation of the three parameters $\beta_D$, $\gamma_D$, and $\delta$. It's still an important result, especially because it differs significantly from the result reported previously from analysis of the optical RESIS study [1]. The precision of the present result is largely due to the precision of the fitted parameter $B_6$, but in also includes an assumed 10% uncertainty in the parameters $\beta_D$ and $\gamma_D$ and the ± 30 a.u. uncertainty in the parameter $\delta$. Both the present and the previous experimental values are shown in Table III. The primary reason for the difference is that the optical data in reference [1] was parameterized by a linear fit. This was sufficient to account for the less precise optical data, but it gave a much larger value of $B_6$ (13(2)). Analysis of the optical data also neglected the contributions of $\gamma_D$ and $\delta$, but this had much less effect on the inferred value of $\alpha_Q$. The value of $\alpha_Q$ reported here is in fair agreement with calculations shown in Table III, although the agreement is not improved going from the uncorrelated Dirac Hartree Fock (DHF) to the presumably more accurate Relativistic Random Phase Approximation (RRPA) calculation. The 18% precision of the present result for $\alpha_Q$ could be improved if the precise rf data pattern could be extended to lower $L$ states to more precisely define the curvature of the polarization plot, or if the individual fine structure intervals could be measured with higher precision. It would also be helpful to have actual calculations of the parameters $\gamma_D$ and $\delta$ to replace the crude estimates made here.

## CONCLUSIONS

The dipole and quadrupole polarizabilities of Rn-like $Th^{4+}$ have been measured through spectroscopy of high-L Rydberg levels of $Th^{3+}$. The new value of $\alpha_D$ is an order of magnitude more precise than the previous measurement [1]. The new value of $\alpha_Q$ differs significantly from a previous report due to the influence of higher order terms in the effective potential describing the interaction between Rydberg electron and $Th^{4+}$ core. Both measured polarizabilities represent demanding tests of *a-priori* theoretical methods used to describe the Rn-like $Th^{4+}$ ion.

## ACKNOWLEDGEMENTS

The work reported here was carried out in the J.R. Macdonald Laboratory of Kansas State University. We are grateful for the cooperation of the laboratory staff and management. We would also like to thank both Donald Beck and Marianna Safronova for providing their unpublished theoretical calculations. The work was supported by the Chemical Sciences, Geosciences, and Biosciences Division of the Office of Basic Energy Science, U.S. Department of Energy.

## APPENDIX A

The core properties that appear in the full effective potential in the text are explicitly defined in term of matrix elements and excitation energies of the free $Th^{4+}$ ion.

$$\alpha_D \equiv \frac{2}{3} \sum_{\lambda'} \frac{\left\langle gJ=0 \|\vec{D}\| \lambda'J=1 \right\rangle^2}{\Delta E(\lambda')}$$

$$\alpha_Q \equiv \frac{2}{5} \sum_{\lambda'} \frac{\left\langle gJ=0 \|\vec{Q}\| \lambda'J=2 \right\rangle^2}{\Delta E(\lambda')}$$

$$\beta_D \equiv \frac{1}{3} \sum_{\lambda'} \frac{\left\langle gJ=0 \|\vec{D}\| \lambda'J=1 \right\rangle^2}{\Delta E(\lambda')^2}$$

$$\gamma_D \equiv \frac{1}{6} \sum_{\lambda'} \frac{\left\langle gJ=0 \|\vec{D}\| \lambda'J=1 \right\rangle^2}{\Delta E(\lambda')^3}$$

$$\delta \equiv \frac{4\sqrt{2}}{15} \sum_{\lambda',\lambda''} \frac{\left\langle gJ=0 \|\vec{D}\| \lambda'J=1 \right\rangle \left\langle \lambda'J=1 \|\vec{D}\| \lambda''J=2 \right\rangle \left\langle \lambda''J=2 \|\vec{Q}\| gJ=0 \right\rangle}{\Delta E(\lambda')\Delta E(\lambda'')}$$

$$+ \frac{2\sqrt{30}}{45} \sum_{\lambda',\lambda''} \frac{\left\langle gJ=0 \|\vec{D}\| \lambda'J=1 \right\rangle \left\langle \lambda'J=1 \|\vec{Q}\| \lambda''J=1 \right\rangle \left\langle \lambda''J=1 \|\vec{D}\| gJ=0 \right\rangle}{\Delta E(\lambda')\Delta E(\lambda'')}$$

$J$ denotes the angular momentum of the state and the sum being over all possible excited states. The dipole and quadrupole operators are defined by

$$\vec{D} \equiv \sum_{i=1}^{86} r_i C^{[1]}(\hat{r}_i) \quad \vec{Q} \equiv \sum_{i=1}^{86} r_i^2 C^{[2]}(\hat{r}_i)$$

Figure 1. Diagram of the microwave RESIS apparatus. A beam of the $Th^{4+}$ is produced using an ECR and charge and mass selected at (1). The beam then passes through a dense Rb target at (2) where it charge captures a highly excited Rydberg electron to become a beam of $Th^{3+}$ Rydberg states. The beams are then charge analyzed at (3) to select just the $Th^{3+}$ Rydberg beam. An electrostatic lens then focuses that beam and ionizes any weakly bound Rydberg states, (4). At (5) a Doppler tuned $CO_2$ laser excites transitions between a specific $nL$ level to the higher $n'$ level, depleting the population in the $nL$ level. Then the rf region, if on resonance with a transitions, will repopulate the $nL$ level at (6). A second Doppler tuned $CO_2$ laser then excites the population of the $nL$ level again the to the higher $n'$ level at (7). The $n'$ level is then Stark ionized and deflected into the channel electron multiplier at (8).

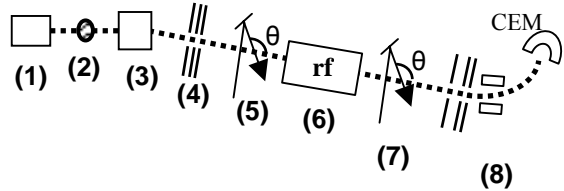

Figure 2. The figure illustrates a RESIS rf signal representing the transition between the L=11 and L=12 levels of the n =37 state of $Th^{3+}$. The rf electric field is propagating anti-parallel to the ion beam velocity. Each data point represents 2 minutes of data taking. The width of the line is due to the transit time through the rf region and the unresolved spin splitting.

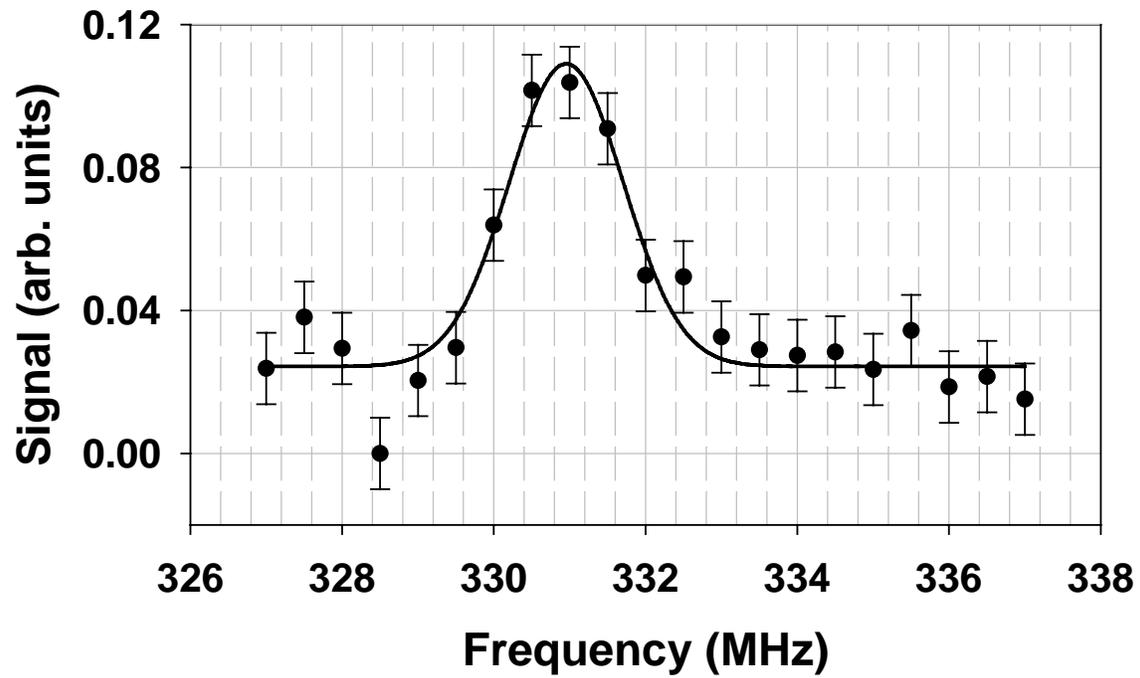

Figure 3. Plot of the scaled first order energies measured in this study. The data is the solid black points with error bar; on some points the error bar lies inside the point itself. The dotted line is the linear fit of the data that clearly fails to account for the measurements. The solid line is a fit that includes terms proportional to $\langle r^{-8} \rangle$, as described in the text. This second fit matches the data within the measurement precision and gives the y-intercept shown by the open circle.

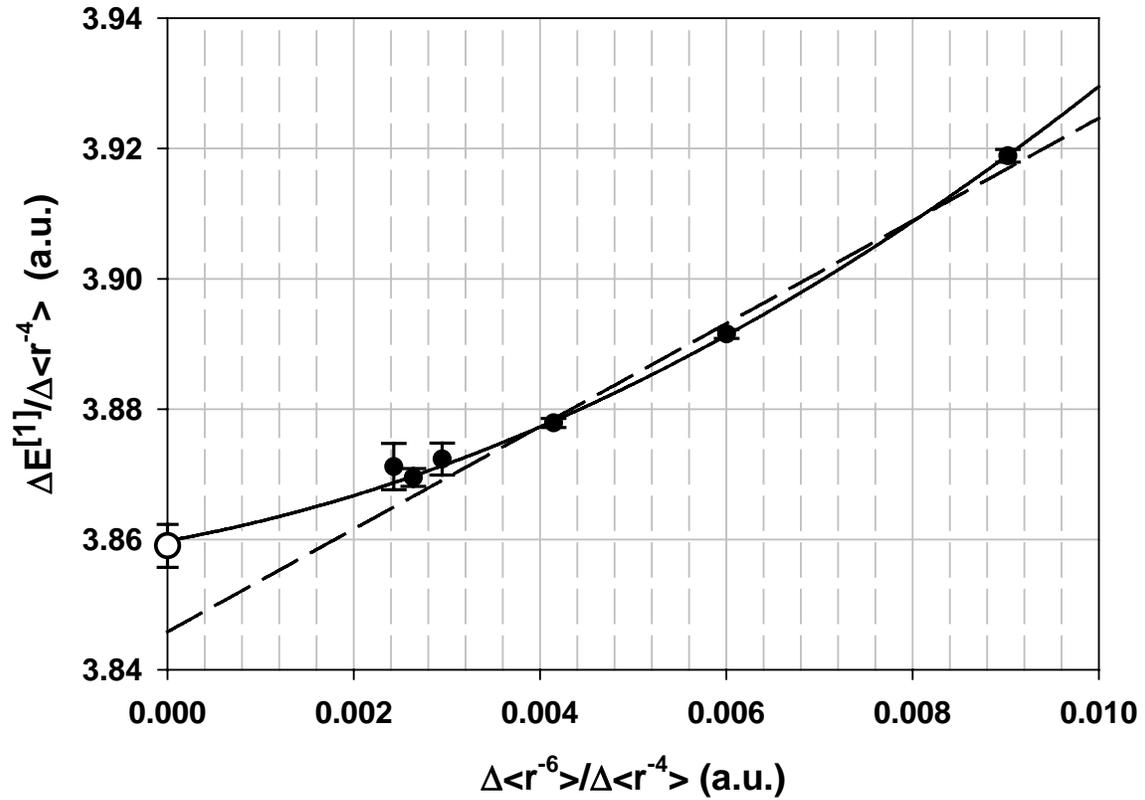

Table I. Measured fine structure intervals in $n=37$ of $Th^{3+}$. Column 1 identifies the transitions, listing the two $L$'s whose separation is measured by their numerical value. Column 2 gives the standard identification using spectroscopy notation, although this becomes awkward for such high $L$s. Column 3 shows the number of independent observations, column 4 shows the net correction due to AC shifts of the resonance position, and column 5 gives the measured interval.

| Interval ($L$-$L'$) | | # Observations | $\Delta f_{AC}$ | $f_0$(MHz) |
|---|---|---|---|---|
| 9-10 | M-N | 2 | 0 | 1008.63(25) |
| 10-11 | N-O | 6 | 0 | 562.08(10) |
| 11-12 | O-Q | 6 | 0 | 331.34(6) |
| 12-13 | Q-R | 5 | 0 | 204.60(13) |
| 12-14 | Q-T | 5 | -0.16(6) | 335.78(12) |
| 12-15 | Q-U | 4 | -0.66(36) | 423.15(39) |

Table II. Corrections applied to infer the portion of the measured intervals due to the expectation value of $V_{eff}$. Column 1 lists the interval, column 2 shows the calculated relativistic contribution to the interval, column 3 shows the calculated second order contribution of $V_{eff}$, including only the leading term, and column 4 shows the corrected interval. For reference, column 5 lists the DC Stark shift rate of each interval.

| Interval($L$-$L'$) | $\Delta E_{rel}$ | $\Delta E^{[2]}$ | $\Delta E^{[1]}$ | $\kappa$ (MHz/(V/cm)$^2$) |
|---|---|---|---|---|
| 9-10 | 8.88 | 1.90 | 997.85(25) | -20 |
| 10-11 | 7.33 | 0.59 | 554.16(10) | -28 |
| 11-12 | 6.16 | 0.20 | 324.98(6) | -38 |
| 12-13 | 5.25 | 0.08 | 199.27(13) | -49 |
| 12-14 | 9.77 | 0.11 | 325.90(12) | -109 |
| 12-15 | 13.71 | 0.12 | 409.32(39) | -183 |

Table III. Comparison of measured and calculated polarizabilities of Th$^{4+}$. All results in atomic units. The experimental results for $\alpha_Q$ rely on calculated values of $\beta_D$, $\gamma_D$ and $\delta$, as described in the text.

|  | $\alpha_D$(a.u.) | $\alpha_Q$(a.u.) |
|---|---|---|
| Experiment | 7.720(7)[a] | 21.5(3.9)[a] |
|  | 7.61(6)[b] | 47(11)[b] |
|  |  |  |
| Theory | 7.699[c] |  |
|  | 7.75[d] | 28.8[e] |
|  | 8.96[f] | 24.5[f] |
|  | 10.26[g] |  |

[a] This work
[b] Reference [1]
[c] RCCSD(T), Schwerdtfeger and Borschevsky, reference [12]
[d] RRPA, Safronova, reference [9]
[e] RRPA, Safronova, reference [10]
[f] DHF, Safronova, reference [10]
[g] Fraga, Karwowski, and Saxena, reference [13]